 \documentclass[final,5p,times,twocolumn]{elsarticle}

\usepackage{amssymb}

\journal{Physica C}

\begin{document}

\newcommand{\con}{cond-mat/}
\newcommand{\eli}{$\acute{{\rm E}}$liashberg }
\newcommand{\vi}{\vec{r}}
\newcommand{\vj}{\vec{r'}}
\renewcommand{\k}{\vec{k}}
\newcommand{\kk}{\vec{k'}}
\newcommand{\q}{\vec{q}}
\newcommand{\Q}{\vec{Q}}
\newcommand{\e}{\varepsilon}
\newcommand{\ee}{\varepsilon^{'}}
\newcommand{\s}{{\mit{\it \Sigma}}}
\newcommand{\J}{\mbox{\boldmath$J$}}
\newcommand{\vv}{\mbox{\boldmath$v$}}
\newcommand{\Jh}{J_{{\rm H}}}
\newcommand{\LL}{\mbox{\boldmath$L$}}
\renewcommand{\SS}{\mbox{\boldmath$S$}}
\newcommand{\Tc}{$T_{\rm c}$ }
\newcommand{\Tcf}{$T_{\rm c}$}
\newcommand{\etal}{{\it et al.}, }
\newcommand{\PRL}{Phys. Rev. Lett. } 
\newcommand{\PRB}{Phys. Rev. B } 
\newcommand{\JPSJ}{J. Phys. Soc. Jpn. } 
\newcommand{\vr}{\vec{r}}
\newcommand{\vrr}{\vec{r'}}
\newcommand{\vrrr}{\vec{r}_{2}}
\newcommand{\vrrrr}{\vec{r}_{3}}

\begin{frontmatter}



\title{Angular Fulde-Ferrell-Larkin-Ovchinnikov superfluid due to 
self-one-dimensionalization in imbalanced cold fermion gases}


\author{Youichi Yanase}

\address{Department of Physics, University of Tokyo, Tokyo 113-0033, Japan 
\\ and Department of Physics, Niigata University, Niigata 950-2181, Japan}

\begin{abstract}
  We study the angular Fulde-Ferrell-Larkin-Ovchinnikov 
(FFLO) state, in which the rotation symmetry is spontaneously broken, 
in population imbalanced fermion gases. 
 The superfluid gases at near $T=0$ are investigated 
on the basis of the Bogoliubov-de Gennes (BdG) equation. 
 We find that the angular FFLO state is stabilized in the gases 
confined in the toroidal trap, but not in the harmonic trap. 
 We discuss the mechanism of the angular FFLO state based on the 
self-one-dimensionalization of the superfluid gas. 
\end{abstract}

\begin{keyword}
\PACS 71.10.Ca \sep 03.75.Hh \sep 03.75.Ss \sep 05.30.Fk 
\end{keyword}

\end{frontmatter}



\section{Introduction}

 Cold fermion gases provide vast opportunities to 
study novel quantum condensed states \cite{giorginireview}. 
 One of the goals of current studies is the realization of 
the FFLO state \cite{FF,LO} in population imbalanced superfluid gases 
\cite{Zwierlein01272006,Partridge01272006}. 
 However, no experimental evidence has been obtained for the FFLO state 
in cold fermion gases. 
 This is mainly because the cold atom gases lack the translation symmetry 
owing to the trap potential, and therefore the spontaneous breaking 
of the space symmetry has never been observed. 
 Then, it is difficult to differentiate the FFLO state from the 
phase separation \cite{mizushima2007}.

 Although the translation symmetry is absent in cold fermion gases, 
the rotation symmetry can be present. 
 Therefore, it is highly desired to produce and study the FFLO state 
with broken rotation symmetry. 
 For this purpose, we have studied the angular FFLO (A-FFLO) state 
in which the rotation symmetry is spontaneously broken 
\cite{yanasecoldatom,shimahara}. 
 We proposed that the A-FFLO state is realized in the toroidal trap, 
although the harmonic trap cannot produce the A-FFLO state. 
 In other words, the FFLO state has not been observed in cold fermion gases, 
because the experiments were carried out by using the harmonic trap. 
 Therefore, it is highly desired to investigate the superfluidity 
in the toroidal trap. 
 In this paper, we clarify the mechanism of the A-FFLO state in 
the toroidal trap, and discuss why the A-FFLO state is not 
realized in the harmonic trap.

\section{Model and Results}

 To discuss the imbalanced fermion gases in the trap, we adopt 
the two-dimensional lattice Hamiltonian given as 
\begin{eqnarray}
\label{eq:attractive-Hubbard-model}
&& \hspace*{-10mm}
H= -t \sum_{<\vi,\vj>,\sigma} c_{\vi,\sigma}^{\dag}c_{\vj,\sigma} 
+ \sum_{\vi \sigma} (V(|\vi-\vi_{0}|) - \mu_{\sigma}) \hspace{0.5mm} n_{\vi,\sigma}
\nonumber \\ && \hspace*{0mm}
+ U \sum_{\vi} n_{\vi,1} \hspace{0.5mm} n_{\vi,2}, 
\end{eqnarray}
where $\sigma = 1,2$ denote two hyperfine states, 
$\vi_{0}$ is the center of the trap, and 
$n_{\vi,\sigma} = c_{\vi,\sigma}^{\dag}c_{\vi,\sigma}$ 
is the number operator of $\sigma$ particles. 
 The two-dimensional gas is produced by the 
one-dimensional optical lattice along the axial direction as well as by the 
pancake potential $\omega_{\rm z} \gg \omega_{\perp}$ with  
$\omega_{\rm z}$ and $\omega_{\perp}$ being the harmonic trap frequency 
along the axial and radial direction, respectively. 
 We take the unit $\hbar = c = 1$. 
 The symbol $<\vi,\vj>$ denotes the summation over nearest neighbour sites. 
 The chemical potential $\mu_{\sigma}$ for $\sigma$ particles 
is determined so that the particle number of each state is $N_{\sigma}$.  
 The particle number and the imbalance are expressed as 
$N = N_{1} + N_{2}$ and $P=(N_{1} - N_{2})/(N_{1} + N_{2})$, respectively. 
 The lattice model is adopted here for simplicity, but we have confirmed that 
the discreteness of the lattice is negligible by assuming the 
small particle density $N/N_{\rm L} \leq 0.1$, 
where $N_{\rm L} = L \times L$ is the number of lattice sites. 
 Therefore, the following results are valid for continuous systems 
without lattices in the two-dimensional space. 
 We take the unit of length  $d$ so that $1/2 m d^{2}=t=1$, where 
$m$ is the mass of atoms. 
 The last term of eq.~(1) describes the $s$-wave attractive interaction. 
 We assume $U/t=-5$ throughout this paper.

 The trap potential is assumed to be 
$V(r)= \frac{1}{2} \omega_{\rm ho} (r/r_{0})^{2} 
+ \omega_{\rm tr} \exp(-r/\xi) $ with $\xi=5$.  
 This potential describes the harmonic trap for $\omega_{\rm tr}=0$ 
and the toroidal trap for $\omega_{\rm tr} \ne 0$. 
 We found that the A-FFLO state is stabilized for 
any $\omega_{\rm tr}/\omega_{\rm ho} > 0$, whose reason will be discussed later. 
 We here show the results for the toroidally trapped system with 
$\omega_{\rm tr}/\omega_{\rm ho} = \frac{2}{3}$ and 
$\omega_{\rm tr}/\omega_{\rm ho} = \frac{1}{4}$
and for the harmonically trapped system with 
$\omega_{\rm tr}/\omega_{\rm ho} = 0$.

 We analyze the model on the basis of the mean field BdG equation. 
Details of the calculation have been shown in Ref.~7. 
 It has been shown that the effects of the fluctuation beyond the BdG equation 
do not alter the qualitative results \cite{yanasecoldatom}. 
 
 Figure~1 shows the spatial dependence of the superfluid order parameter 
for various trap potential and population imbalance. 
 Figures.~1(a-c) show the results for 
$\omega_{\rm tr}/\omega_{\rm ho} = \frac{2}{3}$. 
Because the fermion atoms are accumulated in the highly toroidal trap, 
the superfluid order parameter shows a toroidal shape 
for $P=0$ (Fig~1(a)). With increasing the imbalance, the order parameter 
changes the sign along the radial direction as shown in Fig.~1(b), but 
no spatial symmetry is broken. 
 This is called the radial FFLO (R-FFLO) state. 
 The A-FFLO state is stabilized for $P > 0.43$, and its typical 
spatial structure is shown in Fig.~1(c). 
 It is clearly shown that the rotation symmetry is spontaneously broken 
in the A-FFLO state. 
 We calculated the particle density and the local population imbalances of 
the A-FFLO state, and pointed out several intriguing spatial structures 
which may be observed in future experiments \cite{yanasecoldatom}.

\begin{figure}
\hspace*{-5mm}
\includegraphics[width=9.5cm]{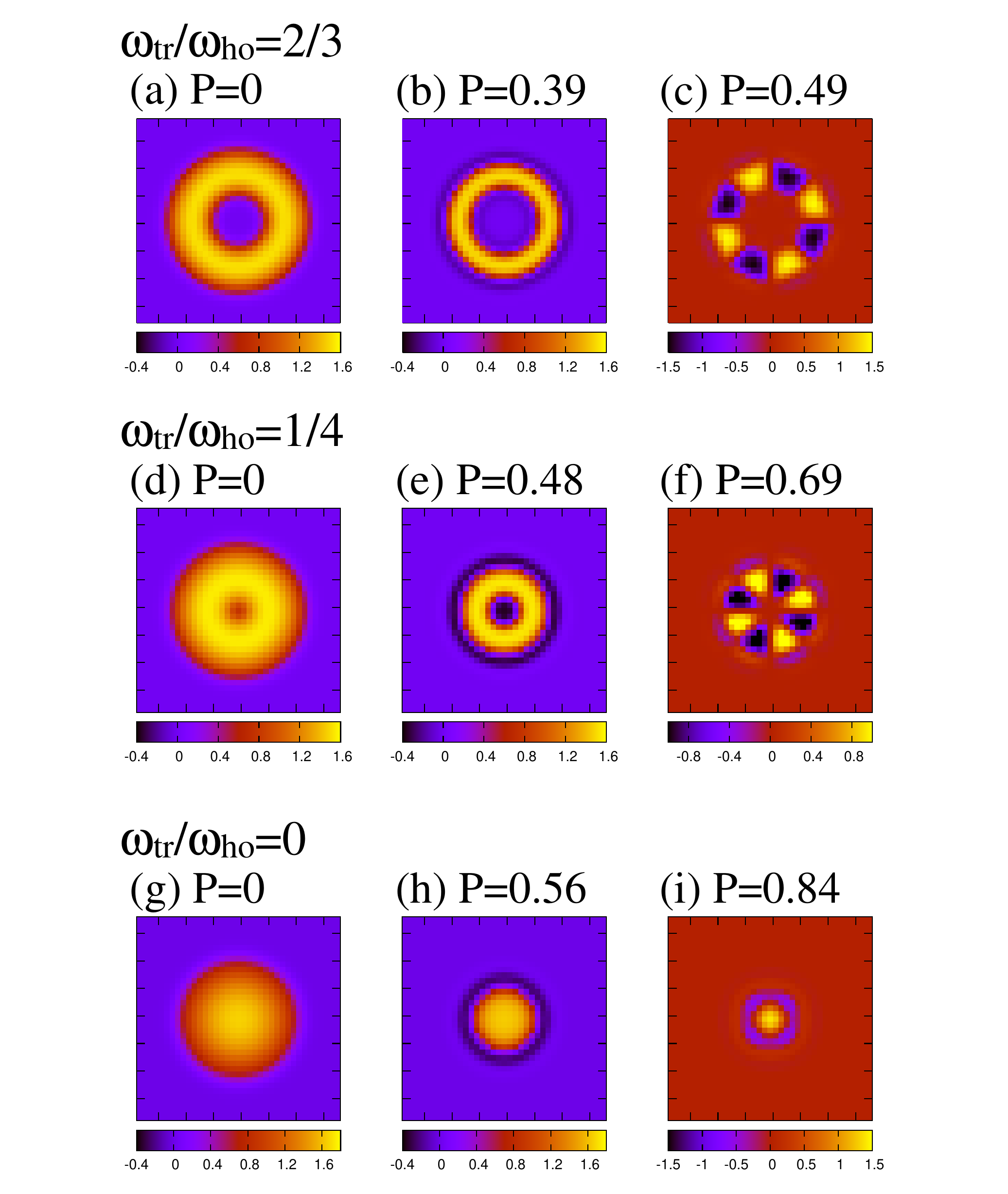}
\caption{
Spatial dependence of the superfluid order parameter 
$\Delta(\vr) = U <c_{\vi,1} c_{\vi,2}>$ 
at $T/t=0.001$, where $T$ is the temperature. 
We assume $\omega_{\rm tr}/\omega_{\rm ho}=2/3$,  
$\omega_{\rm tr}/\omega_{\rm ho}=1/4$,  
and $\omega_{\rm tr}/\omega_{\rm ho}=0$,  
in (a-c), (d-f), and (g-i), respectively. 
The population imbalance is shown in the figure. 
(a), (d), and (g) show the BCS state, while (b), (e), (h), and (i) show 
the R-FFLO state. The rotation symmetry is conserved in these states, 
but broken in the A-FFLO state in (c) and (f). 
See the text for details. 
We assume $N/N_{\rm L} = 0.1$ in (a-f) and $N/N_{\rm L} = 0.07$ in (g-i). 
}
\end{figure}

 The A-FFLO state is also realized for a weakly toroidal trap 
$\omega_{\rm tr}/\omega_{\rm ho} = \frac{1}{4}$, as shown in Figs.~1(d-f). 
 We do not see the clear toroidal structure of superfluid order parameter 
at $P=0$ (Fig.~1(d)), however, the moderate imbalance $P \sim 0.5$ leads to 
the clear signature of the toroidal trap as shown in Fig.~1(e) where 
the R-FFLO state is realized. 
In Fig.~1(e), the sign change along the radial direction induces the 
quasi-one-dimensional structure of superfluid order parameter. 
 This quasi-one-dimensional structure is a characteristic feature of the 
toroidally trapped system as shown in Fig.~1(a). 
 Thus, the feature of the toroidal trap becomes obvious in the 
imbalanced gases. 
 Because the FFLO state is stable in the quasi-one-dimensional system 
\cite{machida1984}, the sign change of the order parameter along 
the angular direction occurs as in Figs.~1(c) and 1(f). 
 Thus, the A-FFLO state is an analog of the quasi-one-dimensional 
FFLO state. 
 Even for the system which is not originally quasi-one-dimensional, 
the {\it self-one-dimensionalization} of the superfluid component occurs 
due to the formation of the R-FFLO state, and therefore the A-FFLO state 
is stabilized in the highly imbalanced gases. 
 Since this mechanism is relevant even for a significantly weak toroidal trap 
$\omega_{\rm tr}/\omega_{\rm ho} > 0$, the presence of the A-FFLO 
state is generally expected in the toroidal trap.  
 
 Because the self-one-dimensionalization discussed above does not occur in 
the harmonic trap with $\omega_{\rm tr}/\omega_{\rm ho} =0$, the 
A-FFLO state is not stabilized in the harmonic trap. 
 We stress again that this is the reason why the FFLO state has never been 
observed in the cold fermion gases. 
 Figures.~1(g-i) show that the rotation symmetry is not broken in the harmonic 
trap. 
 We understand this result by considering the distribution of fermion atoms. 
 The particle density is the largest at the center of the harmonic trap. 
 Therefore, the superfluid phase is stable at the trap core, while the 
superfluidity is suppressed around the trap edge by increasing the population 
imbalance. 
 Since the order parameter at the trap core vanishes in the A-FFLO state, 
the A-FFLO state is highly unstable in the harmonic trap. 
 Thus, the harmonic trap does not produce the broken spatial symmetry in 
the imbalanced superfluid gases.

\section{Summary and Discussion}
  
  We have shown that the A-FFLO state is stabilized in the population 
imbalanced fermion gases confined in the toroidal trap, but not in the 
harmonic trap. 
 The formation of the R-FFLO state leads to the  
self-one-dimensionalization of the superfluid gas and stabilizes 
the A-FFLO state in the highly imbalanced gases. 
 The search of the FFLO state in cold fermion gases has been fruitless 
probably because the experiments were carried out for the harmonically 
trapped gases. 
 It is difficult to detect the FFLO state in the harmonic trap since 
no space symmetry breaking occurs. 
 On the other hand, the rotation symmetry is spontaneously broken 
in the A-FFLO state.  
 We suggest that the experiment in the toroidal trap will realize 
the FFLO state with a broken rotation symmetry 
and will obtain the unambiguous evidence for the FFLO state 
which has been searched for more than 40 years after the theoretical 
predictions \cite{FF,LO}.


\begin{thebibliography}{00}

\bibitem{giorginireview}
S. Giorgini, L. P. Pitaevskii, and S. Stringari, Rev. Mod. Phys. 
{\bf 80} (2008) 1215. 


\bibitem{FF}
P. Fulde and R. A. Ferrell, Phys. Rev. {\bf 135} (1964) A550.


\bibitem{LO}
A. I. Larkin and Yu. N. Ovchinnikov, Zh. Eksp. Teor. Fiz. {\bf 47} (1964) 
1136 [Sov. Phys. JETP {\bf 20} (1965) 762.] 

\bibitem{Zwierlein01272006}
M. W. Zwierlein \etal 
Science {\bf 311} (2006) 492. 

\bibitem{Partridge01272006}
G. B. Partridge \etal
Science {\bf 311} (2006) 503. 

\bibitem{mizushima2007}
T. Mizushima, M. Ichioka, and K. Machida, 
J. Phys. Soc. Jpn. {\bf 76} (2007) 104006. 

\bibitem{yanasecoldatom}
Y. Yanase, Phys. Rev. B {\bf 80} (2009) 220510(R). 


\bibitem{shimahara}
For a work on the FFLO state with rotation symmetry,  
H.Shimahara, Czech. J. Phys. {\bf 46} (1996) Suppl.S2, 561. 


\bibitem{machida1984}
Y. Matsuda and H. Shimahara, 
J. Phys. Soc. Jpn. {\bf 76} (2007) 051005 and 
references there in. 

\end{thebibliography}
\end{document}